# Light Propagation in Linear Arrays of Spherical Particles

**Michael Gozman,** [1, 3] **Ilya Polishchuk,** [1, 2, 3*] **Alexander Burin**[1*]

[1]*Department of Chemistry, Tulane University, New Orleans*

[2]*Max Planck-Institute für Physik Komplexer Systeme, D-01187 Dresden, Germany*

[3]*RRC Kurchatov Institute, Kurchatov Sq 1, 123182 Moscow, Russia*

[*]*Corresponding authors:aburin@tulane.edu*

A propagation of dipolar radiation in a *finite* length linear chain of identical dielectric spheres is investigated using the multisphere Mie scattering formalism (MSMS). A frequency pass band is shown to be formed near every Mie resonances inherent in the spheres. The manifestation of the pass band depends on the polarization of the travelling radiation. To prove this effect, a point dipole placed by the end of the chain is used as an external source of radiation. It is found that, if this dipole is directed parallel to the to the chain axis, the frequency pass bands exist if the refractive index of dielectric spheres is sufficiently large $n_r > 1.9$. For the dipole normal to the chain axis, the pass band can *always* be formed if the chain is sufficiently long. Such a distinction is due to different behavior of the far-field dipolar interaction between the spheres induced by the external source. The edges of the pass bands are defined by the guiding wave criterion based on the light-cone constraint. The criterion of creation of the pass bands correlate with condition of formation of high quality factor modes in these systems found in our previous papers. A comparison with the results available for infinite chains is made. In particular, we clarify the nature of braking down the band structure for small enough wavevectors.

## INTRODUCTION

One-dimensional periodic chains of *metallic* nanospheres can be used to manipulate the optical energy at subwavelength range. In particular, it has been experimentally demonstrated in [1] and numerically justified in [2] that optical pulse can propagate along linear chains of spherical noble nanoparticles. The Ohmic loss inherent in metals is a crucial obstacle to use the metallic nanochains as waveguides. Recently it has been proposed to use one-dimensional periodic chains of *dielectric* micro and nanospheres to fabricate high quality resonators, effective waveguides or microlasers [3-6]. The *infinite* dielectric chains have been theoretically investigated in many papers by means of calculating their band structures (see, ex. Refs [6, 7]). T Here we investigate the optical properties of the *finite* dielectric chains since this case is more relevant to experiment. It occurs that the ends of the chains influence the properties of the pass bands.

We consider a linear array of dielectric spheres centered in the points with coordinates $a, 2a, 3a, ...., Na$, $N$ is the total number of spheres in the chain. In what follows, we consider the spheres touching each other. Such a choice is particularly interesting because in this case the wave-guiding criterion (see below Eq. (4)) is satisfied most easily [3,4]. Thus, we analyze the case when the diameter of the spheres $d = a$.

Energy transport in dielectric chains is due to propagating polariton modes. To investigate the properties of these modes, let us place a point oscillating dipole onto the chain axis at the zero coordinate point. The point dipolar source is considered because it is most strongly connected to the modes of the chain through evanescent field. Practically, this source can be created by placing a die molecule inside a particle [5].

In our previous works [3, 4] the multisphere Mie scattering formalism MSMS formalism [3,4,8 -9] was utilized to investigate the behavior of the quality factor of guiding modes in linear and circular arrays of dielectric (lossless) spheres. In the present paper, using this formalism, we investigate how the dipole radiation propagates along the linear chain and discuss our findings in the light of previous works [3, 4, 7]. For this purpose we calculate how the field amplitude near each sphere depends on the number of a sphere and the frequency of the oscillating dipole. We investigate the cases of this dipole directed either normally or parallel to the chain. Also both magnetic and electric dipolar radiation are considered. The results for the finite chains obtained in this paper are compared with those available in the literature for infinite chain. The most interesting feature of the pass bands is that its low edge coincides with the frequency where the band structure of the corresponding infinite chain breaks down. For the first time a similar feature of the exciton spectrum in low-dimensional molecular crystals has been pointed out to in Ref. [10]. The breaking down of the band structure of infinite chain of dielectric spheres has been established recently in paper [7]. We treat this feature of the polariton spectrum in terms of guiding mode criterion [3, 4] based on the light-cone constraint and consider its relation to the criterion found in Ref. [10].

The dipole approximation used in this paper takes into account only the dipole moments of the dielectric spheres in the linear array under consideration. This approximation has been demonstrated to describe the low frequency guiding modes even quantitatively [3,4]. The paper is organized as follows. We describe briefly the MSMS formalism and give the principle equations describing the dipolar travelling waves which are polariton excitation. Then, using the MatLab software we investigate the condition under which the pass bands are created. In conclusions we summarize the results obtained and consider their physical sense.

## The Multisphere Mie Scattering Formalism

The MSMS formalism was proposed in Refs. [8-10] to investigate scattering and propagation of electromagnetic waves in systems containing spherical particles. In papers [3, 4] we advanced this formalism for studying the quality factor of guiding modes in linear and circular arrays of dielectric spheres. These modes were treated as quasi-states of the system. It this paper we investigate how electromagnetic radiation propagates due to these quasi-states. The MSMS formalism uses the vector spherical wave function (VSWF) as a basis for expansion of solutions to the corresponding Maxwell's equations in the frequency domain. Within this formalism they introduce the VSWF partial amplitude for the incident electromagnetic wave. Let in the dipole approximation they be $p^{l}_{m1}$ and $q^{l}_{m1}$. Here the superscript $l$ points that the basis

VSWFs are associated with the center of $l-th$ sphere, $m=0,\pm1$ describes the polarization of the dipolar radiation defined by the orientation of the active point dipole.

Let $a^l{}_{m1}$ and $b^l{}_{m1}$ be the partial amplitude for the wave scattered by the $l-th$ sphere. These amplitudes obey the system of equation

$$\frac{a_{m1}{}^l}{\bar{a}_1}+\sum_{j\neq l}^{(1,N)}(A^{jl}{}_{m1m1}a_{m1}{}^l+B^{jl}{}_{m1m1}b_{m1}{}^l)=p^l{}_{m1}, \quad (1)$$

$$\frac{b_{m1}{}^l}{\bar{b}_1}+\sum_{j\neq l}^{(1,N)}(B^{jl}{}_{m1m1}a_{m1}{}^l+A^{jl}{}_{m1m1}b_{m1}{}^l)=q^l{}_{m1}. \quad (2)$$

Here $\bar{a}_1,\bar{b}_1$ are the dipole Mie scattering coefficients of the spheres, $A^{jl}{}_{m1m1}$ and $B^{jl}{}_{m1m1}$ are the vector translation coefficients describing the dipole-dipole interaction between $j-th$ and $l-th$ spheres. One can see that, due to axial symmetry of the linear chain, the radiation with different polarization $m$ propagates independently.

If the dipole is parallel to the chain axis, then $m=0$. In this case $B^{jl}{}_{0101}\equiv 0$, and Eqs. (1) and (2) are decoupled. This means that magnetic and electric dipolar radiation propagates independently. In a particular case, when the incident radiation is due to point magnetic dipole directed parallel to the chain axis one has $p^l{}_{m1}=0$, $q^l{}_{m1}=A^{0l}{}_{0101}$. Then, the partial amplitudes for the scattered radiation obey the equation

$$\frac{b_{01}{}^l}{\bar{b}_1}+\sum_{j\neq l}^{(1,N)}A^{jl}{}_{0101}b_{01}{}^l=A^{0l}{}_{0101}. \quad (3)$$

## Propagation of the dipole radiation on the linear chain

We have analyzed the transport properties of the chain of spherical particles with refractive indexes $n_r=3.5,2.7,1.9$ corresponding to $\mathrm{GaAs},\mathrm{TiO_2},\mathrm{ZnO}$. In this paragraph we describe in detail the propagation of *magnetic* dipolar radiation in $\mathrm{GaAs}$. (In particular, this choice is due to the fact that the pass bands corresponding to electric dipole radiation have a larger frequency).

In the case of longitudinal magnetic mode, $m=0$, the polariton propagates according to Eq. (3). This equation has been solved numerically using the MatLab software package. The chain consisting of $N=200$ spheres has been examined. The partial amplitude $b_{01}{}^{l}$ depends on

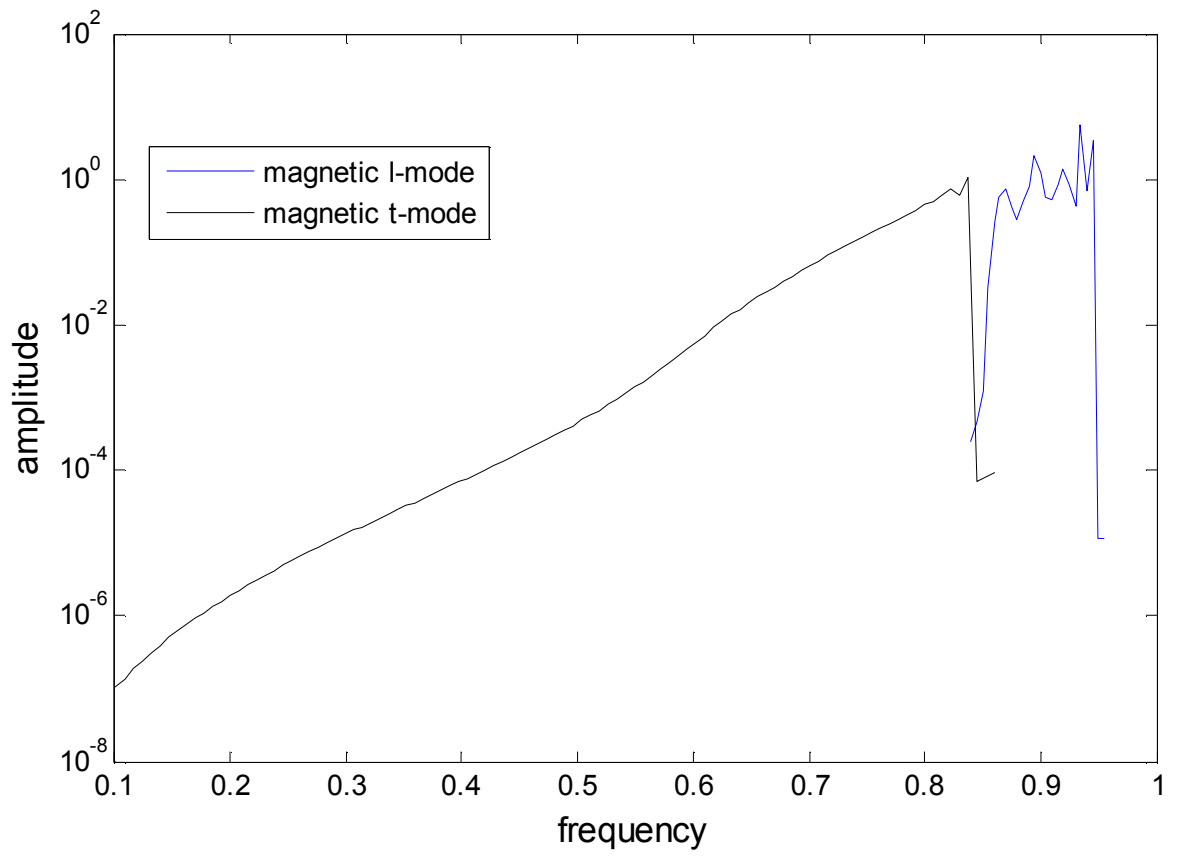

*Fig1. Dependence of the partial amplitude at the 200-th sphere on the dipole radiation frequency.*

the frequency of the incident wave and on the number of a sphere. In Fig.1, the blue curve, the dependence of partial amplitude $b_{01}{}^{200}$ on the frequency is presented for the sphere numbered 200. In this figure the frequency is measured in units $c/2a$, with $c$ and $a$ being the vacuum speeds of light and the diameter of a sphere correspondingly. One can conclude that within the frequency interval $(0.8611,\ 09451)$ this amplitude is practically *a constant,* what points to the pass band formation. This fact has been confirmed also for the cases $N=300, 400$.

To clarify the physical sense of this result let us turn to Fig. 2. In this figure the dependence of partial amplitude on the number of a sphere is presented for different frequencies. One can see that for frequencies out of the pass band interval (a red and a violate curves) the partial amplitude decays with the number of a sphere, while for the frequencies belonging to the pass band the partial amplitude "*oscillates"* with the number of a sphere.

Let us relate the properties of the finite chain with the corresponding infinite chain. In the last case the solution to Eq. (3) is characterized by the quasiwave vector $k<\pi/a$. The travelling polariton modes are characterized by a quasiwave vector $k$. This propagation can be accompanied by an emission of a free photon, which is the principle loss channel. Yet, these losses can be avoided for the polariton modes whose dispersion law obeys the condition

$$\omega(k)\le ck<c\pi/a. \qquad (4)$$

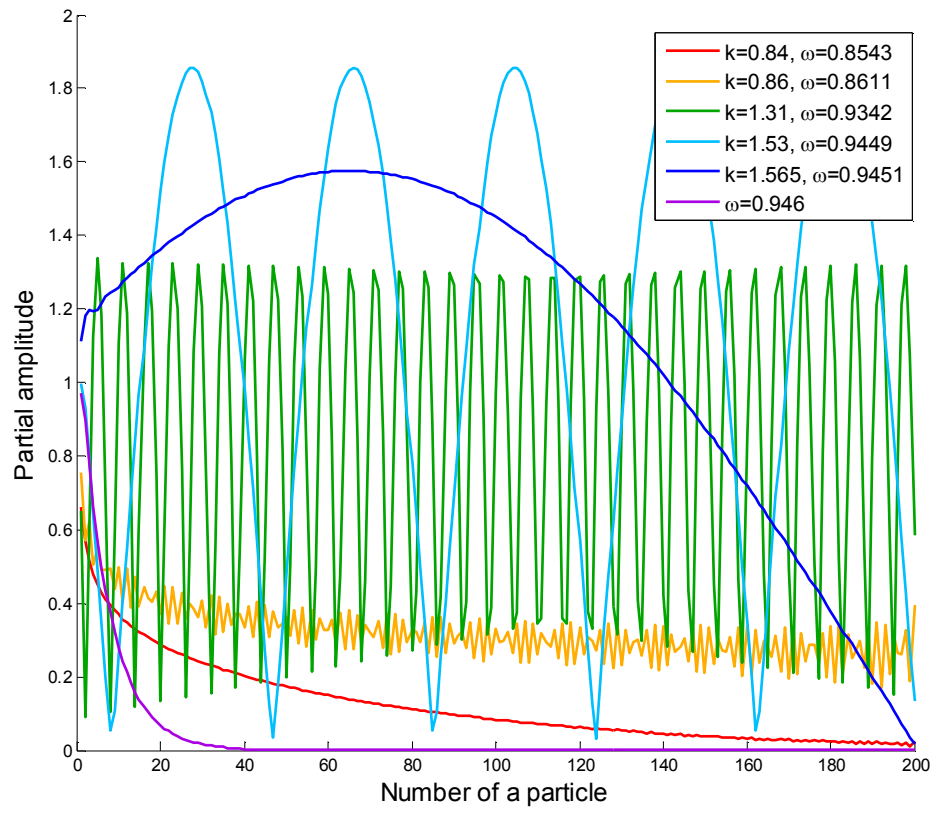

*Fig. 2. Dependence of partial amplitude on the number of a particle for different frequencies*

Guiding mode criterion (4) means that, due to energy and momentum conservation, the transformation of the polariton into a free photon is forbidden because the momentum of emitted photon $\omega / c$ cannot be smaller than its projection $k$ onto the chain axis. It is easy to verify that the polariton modes belonging to the pass band just obey guiding criterion (4). The largest wave vector belonging to the pass band is $k = 1.565 \approx \pi / 2$ and coincides with the top of the Brillouin band of the infinite chain. The low edge of the pass band corresponds to the wave number corresponding to relation $\omega(k) \approx k \approx 0.86$. In the case of the normal dispersion this means that for smaller frequencies the polaritons cannot propagate without decay. Thus, the pass band is formed due to possibility of the radiativeless propagation of a polariton.

It is interesting to note that the similar criterion of radiativeless propagation of excitons in low-dimensional systems has been discovered in Ref. [10]. In this paper a long-wave excitons with $k \ll 1/a$ has been considered. It was established the existence of "acoustic" exciton wave with zero radiative width. Also the authors of Ref. [10] have found the edges of another exciton branch with small radiative width. In the current paper we are interested in only the frequencies, corresponding to the waves with $k \sim \pi / a$. The pass band frequency interval found above is adjacent to the top of the Brillouin band. In spite of the fact that the excitations investigated here and in Ref. [10] are of different nature, they obey the same criterion of propagation without radiation.

Let us also qualitatively explain the oscillating character of the curves in Fig. 2. For the finite chain the edge corresponding to the ending sphere with the number $N = 200$, serves as a *reflector*, resulting to appearance of the extra source. Thus, inside the chain the polariton partial amplitude is an interferention of the *incident* and the *reflected* waves proportional to the expression $e^{ikal} + e^{ika(N-l)}$. The behavior of the curves in Fig. 2 is qualitatively described by this dependence. We have verified that this behavior also does not change at least for $N = 300, 400$. The similar numerical analysis has been made for the chains made of another optical materials, in particular, $TiO_2$ and $ZnO$. We have established that the scenario described above survives both for magnetic and electric longitudinal modes if the refractive index $n_r > 1.9$. Physically this follows from the fact that guiding wave criterion (3) can be fulfilled only if the refractive index is large enough. This circumstance has been pointed out to in Refs [3, 4].

Let us briefly describe the properties of the pass bands corresponding propagation of the *transverse* dipole polaritons in the chain. These modes are characterized by the orbital projection $m = 1$. (The case $m = -1$ is identical to the last one) The result is presented by the black curve in Fig. 1. This curve reveals a sharp upper pass band edge corresponding to the top of Brillouin band of the infinite chain where the guiding criterion is expected to be satisfied most easily. The numerical analysis shows that pass bands for transverse polarization are formed for *any refractive index* if the chain is long enough. This conclusion correlates with the results of Refs.[3,4] where the transverse high quality factor mode has been shown to exist for any dielectric material if the length of the chain was large enough. The low edge of the transverse pass band propagates deeply into the Brillouin band in comparison with longitudinal modes. The distinction between the behavior of the longitudinal and transverse modes is due to different long distance asymptotic behavior of the vector translation coefficients describing the dipole-dipole interaction. These are $1/R$ and $1/R^2$ dependencies for transverse and longitudinal modes, respectively. Thus, the travelling polariton modes do not exist for arbitrary refractive index since the longitudinal mode does not possess the radiation field asymptotic behavior $1/R$.

## CONCLUSIONS

In this paper we studied the propagation of the dipolar radiation along the finite linear chain of dielectric spheres. It is found that the pass bands for the transverse modes always exist if the chain is long enough. One should note that, if the refractive index is small, $n_r < 2$, the length of the chain should be too huge to observe the guidance. At least we cannot see it for the chain of up to *1000* particles. So far as the pass band for longitudinal modes is concerned, it exists only if $n_r > 1.9$. Each frequency pass band is created at frequencies near one of Mie resonances. The upper border of each pass band corresponds to the top of the Brillouin zone of the infinite chain, while the lower border of the pass band is determined by guiding criterion Eq. (4). This means that at small enough wavevectors the frequency spectrum of the quasi-states in the chain becomes unstable with respect to emission of a free photon. The pass bands found above correspond to the high quality factor guiding mode found in [3,4]. Like the guiding modes found in [3,4], the existence of the pass bands is closely connected with validity of guiding criterion (4).

## ACKNOWLEDGEMENTS

This work is supported by the U.S. Air Force Office of Scientific Research (Grant No. FA9550-06-1-0110) and by Russian Fund for Basic Research (Grant No 07-02-00309). We are grateful to Yu. Kagan and V. Babichenko for valuable discussion.